\documentclass[fleqn,12pt,twoside]{article}
\usepackage[headings]{espcrc1}

\readRCS
$Id: espcrc1.tex,v 1.2 2004/02/24 11:22:11 spepping Exp $
\usepackage{graphicx}

\newcommand{\AmS}{{\protect\the\textfont2
  A\kern-.1667em\lower.5ex\hbox{M}\kern-.125emS}}

\title{Quark Matter 2005 -- Theoretical Summary}

\author{Berndt M\"uller
\address{Department of Physics, Duke University, Durham, NC 27708, USA}
}
       
\begin{document}

\maketitle

\begin{abstract}
This is a review of the latest developments in the theory of superdense
nuclear matter, formed in relativistic heavy ion collisions or in the core
of collapsed stars, as they were reported and discussed at the Quark 
Matter 2005 conference in Budapest (Hungary). 
\end{abstract}

\section{INTRODUCTION}

The motivation for the study of superdense matter in relativistic heavy 
ion collisions is the prospect 
of observing a novel state of strongly interacting matter, the quark-gluon 
plasma. Experiments at the CERN Super Proton Synchrotron (SPS) and the 
Relativistic Heavy Ion Collider (RHIC) at Brookhaven have yielded important 
clues of the characteristic signatures of this new state in recent years. 
In fact, the RHIC experiments and Brookhaven National Laboratory (BNL) have
recently announced the discovery of a new form of matter with the properties
of a ``perfect fluid'', i.~e.\ of a strongly interacting plasma-like state
of nuclear matter with an extremely low ratio of shear viscosity to entropy.

The LHC will soon extend the range of energy densities that can be explored 
with measurements of hard probes (jets, heavy quarks, photons). In parallel
to these large-scale experimental programs at major accelerator facilities,
continuing progress in the detection and observation of pulsars holds the 
promise of increasing insight into the limits of stability of neutron stars
and their astrophysical properties. There is good reason for hope that such
studies will ultimately lead to new insights into the equation of nuclear 
matter at low temperature and high baryon density.

It is an understatement to say that this field of research is driven by
experiments at this time. In comparison to the amazing wealth of new and
exciting data from the RHIC and SPS experiments reported at this meeting, 
the progress in our theoretical understanding of the central questions
relating to the properties of quark matter has been modest. This remark
should not be taken to mean that no progress is being made on the theory
front -- I will discuss some important and intriguing new developments below --
but as we learn more about the complexities of the problem, we realize 
that definitive theoretical results will require very large 
investments in time, manpower, and computing resources. All these are 
occurring at the current time, which leads to the expectation that steady
theoretical progress will be made over the next few years.

The role of theory in physics is to develop ideas and concepts which allow
us to quantitatively interpret the experimental data and to propose new 
experimental tests and analysis strategies. We need a steady stream of
new ideas, both ``good'' and ``bad'' ones, because it is notoriously 
difficult to correctly judge the value of a new idea at an early stage
in its development. We also need sustained diligent work on established 
approaches, which have been found to be fruitful and valid, and
we must sort out and safely dispose of failed ideas.
In my discussion of what was presented at this conference I will begin
with the exciting new ideas, I will then report on the progress in the
established areas of theoretical relativistic heavy ion physics and,
finally, I will discuss the status of theoretical insight in various
areas of heavy ion phenomenology. In doing so, I will not be able to
cover every single theory talk at QM2005, and I apologize 
to all those speakers and presenters of posters whose work remains 
unmentioned in this summary.

\section{NEW IDEAS}

\subsection{Thermalization}

One of the greatest puzzles posed by the experimental results from RHIC
has been the apparently almost instantaneous equilibration of the 
produced matter demanded by the flow observables. Perhaps the most 
exciting theoretical development relevant to this issue and reported
at this meeting (by S.~Mrowczy{\'n}ski \cite{Mrow}, M.~Strickland
\cite{Strickland}, and Y.~Nara\cite{Nara}) is the idea that plasma 
instabilities in the soft modes of the gluon field may be responsible 
for the rapid momentum equilibration. The fact that unstable field 
modes exist whenever the particle momentum distribution is anisotropic
has long been known in electromagnetic plasma physics. But only in 
recent years has the potential relevance of these so-called Weibel
instabilities \cite{Weibel} to relativistic heavy ion collisions been 
recognized. 

The basic idea, nicely explained in Mrowczy{\'n}ski's talk, builds 
on the insight that the momentum spectrum of the 
partons, which are deposited by semihard QCD interactions during the
initial impact of the two nuclei, quickly becomes locally anisotropic
due to the longitudinal expansion of the system. Measured in the comoving 
frame, the width of the longitudinal momentum distribution becomes much 
narrower than the width of the transverse momentum distribution:
$\langle \Delta p_L^2 \rangle \ll \langle \Delta p_T^2 \rangle$.
A transverse color current fluctuation with wavelength in the beam
($z$-)direction less than the color screening length in the plasma
will then induce a transverse chromomagnetic field, which tends to 
further amplify the current fluctuation. In a range of modes $k_z$,
this leads to exponential growth of the color field at a rate 
$\Gamma(k_z) \sim k_z < m^*$, where $m^*$ is the effective gluon mass
(proportional to the plasma frequency) in the medium.

Over the past year, the dynamical evolution of this instability has
been studied in detail by means of numerical simulations of
the nonlinear plasma dynamics in the presence of an anisotropic
parton distribution \cite{Arnold,Rebhan,Dumitru}. The numerical studies 
show that the initially exponential growth of the unstable modes 
eventually moderates, when the magnetic energy density in
the soft modes reaches the value $g^2Q_s^4$, where $Q_s$ is the 
typical energy carried by the particles. Thereafter, the energy in
the soft gluon modes grows only linearly, indicating a continuing
cascade of field energy into short wavelength modes. At present, it is
unclear which mechanism triggers the end of the exponential growth.

What is clear, however, is that the strength of the chromomagnetic 
field ($B \sim gQ_s^2$) is sufficient to bend the momentum of the
hard partons on a time scale $t_{\rm iso} \sim p/gB \sim (g^2Q_s)^{-1}$,
which thus governs the isotropization of the parton distribution. Note
that this is an expression which does not involve $\hbar$ and thus
constitutes a {\rm classical} time scale. It is also amusing to note
that $t_{\rm iso}$ is reminiscent of the characteristic growth time of the
(coarse grained) entropy of classical gauge field, which is given by
the Lyapunov exponents $\lambda$: 
$t_s = S/(dS/dt) \sim \lambda^{-1} \sim (g^2 T)^{-1}$ \cite{chaos}.

A detailed dynamical framework of the role of this mechanism in the
process of parton thermalization has yet to be developed. Although it
appears unlikely that the weak-coupling estimate of the time to full
thermalization obtained in the ``bottom-up'' scenario \cite{BMSS} 
will be significantly modified, it is quite likely that the plasma
instabilities can induce a precocious approach to isotropy of the
local momentum distribution. This would already explain the rapid 
transition to a regime, in which fluid dynamics can be applied, as
the comparison of hydrodynamic calculations of the elliptic flow 
$v_2$ with the RHIC data suggests. It may even be possible to devise
an effective description of this early phase in terms of a type of
{\rm chromo-magneto-hydrodynamics}. The bending effect of the soft 
gluon fields would have a similar effect as collisions in limiting 
the rate of momentum transport. Because particle propagation in 
coherent fields conserves entropy, it is not clear whether such an 
effect can lead to an increase in the shear viscosity $\eta$ of the medium. 
Boldly assuming that it does, the general expression for the viscosity, 
$\eta \sim \rho{\bar p}\lambda_f$ (where $\bar p \sim Q_s$ denotes the 
average momentum of a particle), would be replaced with 
$\eta \sim \rho{\bar p}t_{\rm iso} \sim \rho{\bar p}^2/gB \sim \rho/g^2$, 
reducing the shear viscosity by a factor $g^2$ compared with the usual
perturbative result. 

One may ask whether the strong gluon fields 
generated by the Weibel instability can be experimentally observed.
One way may be through the study of event-by-event fluctuations in 
$v_2$ \cite{Mrow-v2}; another way could be via their effect on jets,
which penetrate the plasma in the transverse direction. Since also the
chromomagnetic fields are aligned in the transverse plane, they
can only deflect jets in the longitudinal direction. A rough estimate
suggests that the deflection could be substantial, leading to a 
broadening of the jet cone along the beam direction, but no effect 
on the azimuthal opening angle of the jet. Note that this is exactly 
what is observed experimentally \cite{STAR-jet-cone}.

Another interesting approach to the thermalization is to invoke the
analogy to the Unruh phenomenon of the thermal vacuum in accelerated 
reference frames \cite{Unruh}, presented here by D.~Kharzeev 
\cite{Kharzeev}. As argued by Unruh and first demonstrated in detail
by Fulling \cite{Fulling}, the Minkowski space vacuum of a free 
relativistic quantum field theory appears to a constantly accelerated 
observer exactly like a thermal state. The temperature $T$ and
the acceleration $a$ are related by the formula $T=\hbar a/(2\pi c)$
or $T=a/2\pi$ in natural units. The phenomenon is closely related to,
though more general, than the Hawking effect \cite{Hawking}, by which 
black holes radiate thermally. A constantly accelerated observer also
experiences an event horizon which, in contrast to the case of black
holes, is not a topological property of the space-time geometry.

If this insight could be applied to heavy ion collisions, where strong 
color fields lead to the coherent acceleration of many partons, the 
spontaneous appearance of a nearly thermal system could be neatly 
explained. One might even, with slight hyperbole, call the idea 
``thermalization by black magic''! The question is how far 
the analogy carries. The idea that external fields other than gravity
could be interpreted as mediators of a thermal ensemble has been 
discussed before (see e.~g.~\cite{QED=T}). It turns out that the 
analogy is not perfect, because ordinary electric or chromo-electric
fields do not accelerate all particles equally but differentiate 
between particles with different charges, in particular between 
particles and antiparticles. In other words, quarks and antiquarks 
in a given chromo-electric field do not experience the same event 
horizon, nor do gluons. The equivalence principle only applies to
gravity, not to other interactions. Furthermore, gluon radiation 
can be induced not only by the acceleration of color charges, but 
also by their rotation in SU(3) space. This aspect, which was nicely 
demonstrated in Kovchegov and Rischke's derivation of the gluon 
spectrum emitted in the collision of two sheets of color glass 
condensate \cite{KR97}, is peculiar to QCD and does not occur in QED.
Finally, even if this idea could be invoked with benefit for the 
explanation of a (approximately) thermal initial gluon spectrum, it
would not explain the formation of a system that can be described by fluid
dynamics, because it completely lacks the notion of a mean free path.
Thus, at best, the concepts discussed by Kharzeev could be used to
explain what has recently been called ``pre-thermalization''
\cite{Berges}.

\subsection{Other new ideas}

Let us now turn to other new ideas. I find the idea of jet-induced Mach 
or Cherenkov cones intriguing, because it may help us learn about the 
transport properties of the matter created in heavy ion collisions. But
since this topic was covered in J\"org Ruppert's Focus talk \cite{Ruppert}, 
I refrain from summarizing it here once more. 

Another new idea was discussed by J.~Cramer \cite{Cramer}, who presented 
his and G.~Miller's calculation of identical pion correlations using
the framework of the distorted wave Born approximation (DWBA). Their 
calculation uses the optical potential for pions interacting with a 
dense nuclear medium, which has long been studied in pion-nucleus 
scattering and pionic atoms. The potential has a large imaginary part, 
which describes absorption within the medium. It has been argued that 
such a description is inadequate, because the strongest absorptive 
channels -- such as the $\rho$-meson and the $\Delta$-baryon -- are short-lived
resonances, which mostly decay again by pion emission. In other words,
the absorbed pions often reappear quickly, albeit in different quantum
states. 

However, I believe that this critique is unfounded. The DWBA 
is an approximation to S-matrix theory, and if one includes absorptive
effects in the {\rm final} state, they really describe the time reversed
process, i.~e.\ emission. Since the source term in the HBT formalism 
describes the vertex positions of the last inelastic interactions of
the emitted pions, the DWBA seems to be a valid approach
to a quantum mechanical description that includes the effects of the
in-medium propagation of the observed pions. The one issue that is a 
bit worrisome is that the Cramer-Miller analysis invokes both, a large
pion chemical potential and a strongly attractive potential, which is
interpreted as onset of chiral symmetry restoration. The combination
of these two effects may bring the pion field precariously close to
Bose condensation. It would be good to check whether this is a problem
or not \cite{Cramer-com}. It would also be interesting to apply the 
DWBA approach to other hadrons, such as protons and charged kaons.

As a last example of new developments, which were discussed at this 
meeting, I want to mention the treatment of (soft-hard) recombination 
as a special case of parton fragmentation. Hwa and Yang \cite{Hwa-sh}
, as well as Greco, Ko and Levai \cite{GKL}, have argued 
for some time that this hadronization mode is an important source of
hadrons emitted with momenta $p_T = 2-8$ GeV/c. E.~Wang 
\cite{EWang} showed how soft-hard parton recombination can be formulated 
as a special case of parton fragmentation in the presence of a medium.
This represents an important step toward a unified, QCD-based theory 
of hadron emission at intermediate and high transverse momenta.

\section{ESTABLISHED APPROACHES}

\subsection{Lattice QCD and the equation of state}

One of the best established theoretical approaches in our field is
lattice gauge theory, which is unique in its ability to provide 
model independent predictions for observables beyond
the reach of perturbative QCD. It is no exaggeration to say that 
most of what we believe to know about the structure of hadronic
matter at high temperature is based on results from lattice gauge
theory. Lattice simulations predict that the properties of strongly 
interacting matter without net baryon surplus change drastically at 
an energy density of the order of 1 GeV/fm$^3$, corresponding to a
temperature of about $T_c\approx 170$ MeV. Previous
calculations also indicated that this transition is not a true phase 
transition (at zero net baryon density), but at rapid crossover from 
a low-temperature phase dominated by hadrons into a high-temperature
phase of a plasma of interacting quarks and gluons. The transition 
is seen as a steep rise in the vicinity of $T_c$ in the effective 
number of degrees of freedom $\nu$ defined as 
$s(T)= 2\nu\pi^2 T^3/45$, where $s(T)$ is the entropy density 
as function of temperature.

Up to now, limitations in the available computer power forced lattice 
simulations to make a number of unphysical assumptions. For example,
the quark masses were too large, the quark actions violated chiral
symmetry, and the number of lattice points in the euclidean time 
direction, $N_t$ was usually quite small ($N_t=4$). This required
rather large extrapolations of the numerical results to the physical
point, leading to substantial systematic uncertainties in the 
predictions. The recent advent of computers with multi-teraflops capability 
has changed this. Several calculations of QCD at nonzero temperature
are presently underway with realistic quark masses, finer mesh
sizes, and improved fermion actions. The first, preliminary results
of one of these calculations were reported at this conference by S.~Katz 
\cite{Katz}, who showed a long list of improvements compared to previous
calculations, including physical quark masses, exact Monte-Carlo
algorithms, and $N_t=6$. It may be important to mention, however,
that the calculation does not use one of the best improved fermion
actions, nor a fermion action with an explicit chiral symmetry,
such as domain wall fermions.

The results presented by Katz confirm many of the previous findings.
The transition between the low- and the high-temperature phase is
still a smooth, but rapid crossover. However, the new calculation
seems to suggest a considerably higher value for the critical temperature.
Katz \cite{Katz} reported the values:
$T_c = 186 (3) (3)\, {\rm MeV} (N_t = 4)$ and
$T_c = 193 (6) (3)\, {\rm MeV} (N_t = 6)$ ,
which lie significantly above the values reported at earlier 
Quark Matter conferences 
($T_c^{(N_f=2)} = 173\pm 8\, {\rm MeV}$ and 
$T_c^{(N_f=3)} = 154\pm 8\, {\rm MeV}$ \cite{Karsch02}, see also
\cite{FK04}). If the new values are confirmed, it would force us 
to rethink the message from the chemical equilibrium analyses 
of the hadron yields in relativistic heavy ion collisions. 
The temperature $T_{\rm ch}$ obtained in
these fits is consistently lower than 180 MeV (see e.~g.\ 
Florkowski's talk at this conference \cite{Florkowski}, or 
\cite{chem-fits}). A clear separation between $T_c$ as defined
by the peak in susceptibilities and $T_{\rm ch}$ would imply that
the hadronic system does not chemically freeze out immediately
after the deconfinement transition. The experimentally determined
value of $T_{\rm ch}$ would then not have a universal meaning,
because the chemical freeze-out would not reflect thermodynamic
properties of QCD matter, but aspects of the hadrochemical kinetics,
which is system dependent. This would, of course, not be against
fundamental principles of physics, but it would constitute an 
unfortunate circumstance, because it would make an experimental 
determination of $T_c$ much more difficult. Given these implications, 
it is probably wise to wait for the results of other currently ongoing 
lattice calculations before revising the standard picture.

It is interesting to ask whether it is possible to put data points
on the equation of state curve $\varepsilon(T)$. This would require
the simultaneous measurement of the energy density $\varepsilon$ (or
the entropy density $s$ or the pressure $P$) and the temperature $T$.
The latter may be possible, if thermal photon radiation from the
quark-gluon plasma can be detected \cite{Enterria}. An elegant novel 
experimental method, measuring low-mass
dileptons at high transverse momentum as a proxy, presented here by
the PHENIX collaboration \cite{PHENIX-gammas} is a promising step in this 
direction. As K.~Rajagopal and I have recently pointed out, an alternative
way of determining the thermodynamically active number of degrees of
freedom $\nu$ is to simultaneously measure $\varepsilon$ and $s$ 
\cite{MR05}. We already have a rather precise determination of the
entropy per unit of rapidity ($dS/dy = 5100 \pm 400$, corresponding
to $s = (33 \pm 3)$ fm$^{-3}$ at $\tau = 1$ fm/c). What is needed now is 
an independent determination of the energy density at an early time.
The energy loss of partons may be the best available tool for such
a determination; the report at this meeting of the observation of
the ``punch through'' of the away-side jet \cite{Jacobs} represents
an important progress toward this goal (see also below).

There is steady progress toward a detailed understanding of the 
physics of quark matter at high density, but low temperature. 
K.~Rajagopal \cite{Rajagopal} review the general status of the 
theory of color superconductivity. If the existence of ``gapless'' 
color superconductors (referring to the absence of a gap in some 
pairing channels) is confirmed by improved calculations, it could
have interesting phenomenological consequences in collapsed stars.
If the ground state of the matter has a crystalline structure, of
the ``LOFF'' type, it would allow quark stars to exhibit glitches 
and other phenomena, which are normally thought to require a solid
neutron crust. T.~Sch\"afer \cite{Schafer} discussed the progress
in constructing effective theories for the dynamics near the Fermi
surface when unscreened QCD forces vitiate the formalism of Fermi
liquid theory. On the numerical front, S.~Eijiri \cite{Eijiri} 
presented new results of lattice calculations for nonzero baryon 
chemical potential, using the Taylor expansion around $\mu_B=0$.
The results impressively confirm the validity of the resonance gas 
model (in the baryonic sector) for $T<T_c$. The $\mu_B$-dependence 
of the color screening mass agrees well with perturbation theory for 
$T>1.5T_c$. The calculations have reached a state where many model
predictions can be tested against rigorous predictions of QCD. 

\subsection{Fluid dynamics}

A well established approach to the treatment of the evolution of
the dense matter created in relativistic heavy ion collisions is
fluid dynamics. One of the core insights of the RHIC program is 
that fluid dynamics is successful in describing the collective
flow of hadrons and hadron spectra up to $p_T=2-3$ GeV/c. The strong
rapidity dependence of many observables has made it clear, however,
that fully three-dimensional calculations are needed. Significant
progress on this front was reported at this conference. Y.~Hama
\cite{Hama} presented results from the SPheRIO code, which is uses 
the particle-in-cell method and implements a continuous freeze-out
model, where hadrons are decoupled on the basis of somewhat schematic
mean-free path considerations. T.~Hirano \cite{Hirano} showed results 
from a 3-D Eulerian hydrodynamics code, which allows for deviations 
from chemical equilibrium in the hadronic phase. Hirano, as well as
C.~Nonaka \cite{Nonaka}, whose code is based on Lagrangian fluid 
dynamics, also presented first results from so-called ``hydro+micro'' 
models, in which the evolution above $T_c$ is described by 3-D 
hydrodynamics, and the hadronic phase is treated by a cascade 
simulation. The few glimpses we have seen here raise the expectation
that the new codes will yield a much improved understanding of 
the rapidity dependence of soft observables.

These are important developments for our field, and we can look 
forward with great interest to the results of further systematic 
comparisons of these improved transport models with the experimental 
data, similar to the studies by Hirano and Nara \cite{HiranoNara} of 
the interplay between jet propagation and hydrodynamical evolution. 
It will be highly desirable to see direct comparisons between the 
different codes for the same initial conditions, equation of state,
and other assumptions. The increasing complexity of the various
implementations of what amounts to essentially identical physics
makes it crucially important to establish the validity of these
codes before we use them to draw firm conclusions. (For such a 
study of the performance of the SPHERIO code, see \cite{Aguiar}.)
The immediate goals are the determination from the data of the 
equation of state and the initial conditions reached at RHIC. Other 
important applications will be calculations of the effect of the
medium on hard probes, such as jets, photons, and lepton pairs.

\subsection{Parton transport}

Several speakers reported substantial progress in microscopic 
approaches to parton transport. D.~Teaney's \cite{Teaney} discussed the
momentum evolution of charm quarks in a quark-gluon plasma by
means of a collisional Langevin equation of the form
\begin{equation}
\frac{dp}{dt} = \eta_D p + \xi(t)
\end{equation}
with Gaussian white noise 
$\langle\xi(t)\xi(t')\rangle = (2T^2/3D)\delta(t-t')$ and the
friction coefficient $\eta_D = \frac{T}{MD}$, where $D$ is the diffusion
coefficient of charm and $M$ the charm quark mass. The diffusion
coefficient is related to the shear viscosity $\eta$ by a relation 
of the form 
$D \approx \frac{3}{4\pi\alpha_s T} \approx 6 \frac{\eta}{sT}$.
Teaney showed that the PHENIX data on the suppression coefficient
$R_{AA}$ of non-photonic electrons (presumably originating from
semileptonic charm decays) can be well described by a value
$D = (3\, \cdots 6)/(2\pi T)$,
consistent with a very low value of the shear viscosity 
$\eta/s = (1\, \cdots 2)/(4\pi T)$,
near its conjectured lower bound \cite{Starinets}.

Continuing with the topic of charm transport, R.~Rapp presented
results from a model calculation assuming that charm quarks can
scatter from D-meson like resonances with various hadronic 
quantum numbers in the plasma slightly above $T_c$ \cite{Rapp}.
He and his collaborators (see also the talk by M.~Mannarelli 
\cite{Mannarelli}) find that this mechanism strongly accelerates
the momentum equilibration of $c$-quarks, even for a single 
resonant state and its partners dictated by the heavy quark and 
chiral symmetries. M.~Djordjevic showed
the results of a complete radiative energy loss calculation for
heavy quarks in a quark-gluon plasma \cite{Djordjevic}. On the
basis of a schematic space-time model for the absorbing medium
she concluded that the $R_{AA}$ value for electrons from charm 
decays cannot be lower than twice the $R_{AA}$ value for pions.
This limit is incompatible with the preliminary results from STAR 
and only marginally compatible with the PHENIX data. It would be
good to explore whether the limit holds up in the context of more
realistic models of the space-time evolution. The importance of 
comparing the effects of energy loss on light and heavy quarks 
was also emphasized in N.~Armesto's talk \cite{Armesto}.

There also was news on the front of partonic cascade codes. Z.~Xu
\cite{ZXu} reported results from a new cascade code, which implements
$gg\leftrightarrow gg$ and $gg\leftrightarrow ggg$ scattering
by means of local rates with detailed balance. The inelastic
process turns out to be crucial, as argued by S.~Wong almost
a decade ago \cite{SWong}, not only for chemical but also 
for thermal equilibration. It still needs to be understood in 
more detail, how the $2\leftrightarrow 3$ process manages to
speed up thermalization more than the increased transport cross
section from its more isotropic angular distribution would naively 
suggest. Other implementations of radiative gluon processes 
in parton transport codes did not show a similarly large effect
\cite{Shin}.

Several speakers, such as D.~Molnar \cite{Molnar} and B.~Zhang
\cite{Zhang} presented calculations of binary parton cascades
with hugely cranked up binary cross sections. The studies may 
serve some purpose in schematic explorations of the origin of flow 
and energy loss at the partonic level. However, the question needs 
to be asked what exactly these studies are telling us, beside the 
obvious (that larger partonic cross sections lead to faster equilibration 
and to increased collective flow), and whether their conclusions are
not misleading due to the unrealistic assumptions.

\subsection{Parton saturation}

Another established fruitful idea is parton saturation with
the asymptotic approach to a universal structure, the color 
glass consensate (CGC). K.~Itakura \cite{Itakura} gave an impressive 
review of the recent conceptual and formal developments in this field.
The question whether this asymptotic physics is relevant to the phenomena 
observed at forward rapidities in d+Au collisions at RHIC is still under
discussion (see the talks by G.~Veres \cite{Veres}, D.~R\"ohrich
\cite{Rohrich}, and J.~Jalilian-Marian \cite{Jamal}). The problem is 
that the phenomenology is complex, involving a mixture of initial-state 
and final-state effects. It appears to be difficult to cleanly separate 
different sources of nuclear modifications from each other.

One important new result in the context of the CGC, reported
here by T.~Lappi \cite{Lappi}, is that quark-pair production
in the strong gluons fields generated by the shattering CGC is
quite large. For reasonable values of the parameters, Lappi finds
that several hundred light quark pairs are
produced. On the one hand, this result is welcomed, because it
suggests that a chemically equilibrated quark-gluon plasma is
originally produced. On the other hand, it poses a potential
problem for the standard jet quenching picture, because the 
total number of gluons and quarks is limited by the total entropy
observed at freeze-out. More quarks imply fewer gluons. But
quarks are much less efficient at causing parton energy loss, which
would imply that the slight underestimate of the measured hadron 
suppression by many calculations (see e.~g.\ B.~Cole's lecture
\cite{Cole}) would be further enhanced. It is too early to draw
firm conclusion, but the result suggests that a detailed
analysis of the possible implications for jet quenching of different 
abundance ratios of quarks and gluons in the medium is needed.

\subsection{The strongly coupled quark-gluon plasma}

Practitioners of lattice theory have told us for quite some time
that QCD matter at temperatures not far above $T_c$ is {\em not}
a perturbative quark-gluon plasma. The impressive work which has
been done over the past five years on perturbative resummation 
techniques for the QCD equation of state has confirmed this insight
\cite{BIR,ABS}. The perturbative expansion based on hard-thermal
loop quasiparticles works well for $T>(2\cdots 3)T_c$, but it fails in 
the region near $T_c$, which is explored by the RHIC experiments. If
the plasma below $2T_c$ is a strongly coupled one (a ``sQGP'' in 
the current terminology), the question is whether it contains any 
long-lived quasiparticles and if so, what their structure is.

We may look for guidance to strongly coupled electromagnetic
plasmas, which have been thoroughly studied experimentally and 
theoretically. The physics of these plasmas and its possible 
relevance to our field was superbly reviewed by M.~Thoma 
\cite{Thoma}. The two parameters controlling electromagnetic 
plasmas are temperature and charge carrier density. A line defined
by the condition
$\Gamma_{\rm EM} = q^2/(Ta) = 1$,
where $a$ is the average distance between particles and $q$ their electric
charge, divides the phase diagram into regions of weak ($\Gamma<1$) and 
strong ($\Gamma>1$) coupling. The precise translation of this condition 
to a quark-gluon plasma is not known, but it must roughly have the form
$\Gamma_{\rm QGP} \sim C_2\alpha_s/(Ta) \approx 1$,
where $C_2$ is the eigenvalue of the quadratic Casimir operator for
the color charges making up the plasma. If one inserts numbers, one
finds that $\Gamma_{\rm QGP}$ is, indeed, somewhat larger than unity for
the conditions prevailing in the range $T_c<T<2T_c$. The experience
from electromagnetic physics suggests that a plasma in this range of
parameters is no longer perturbative, but still far away from the
region where regular short-range correlations or even quasi-crystalline
structue are found. Still, the analogy to electromagnetic plasmas is very 
useful, because it suggests phenomena and properties to look for, and maybe
even theoretical tools to apply, in theoretical investigations of the
quark-gluon plasma near the phase boundary.

An conjecture, whose ultimate value is still uncertain, is the dominance
of partonic bound states in the quark-gluon plasma in the region 
near $T_c$, which was discussed by E.~Shuryak \cite{Shuryak}. Such
substructures are difficult to track in lattice simulations, because
they are not color singlets. Their calculation on the lattice, therefore,
requires gauge fixing with all its concomitant pitfalls. It is much
easier, although possibly less conclusive, to look for indirect signals
of the abundant presence of partonic bound states on the lattice. Two
such studies were discussed by V.~Koch \cite{Koch} and F.~Karsch 
\cite{Karsch05} in their talks. Koch and collaborators focus on the 
correlations between strangeness and baryon number.
If all quasiparticles in the plasma have the quantum numbers of 
quarks or gluons, only quarks carry strangeness. If bound states 
of quarks and antiquarks exist as quasiparticles, strangeness is also
carried by states without baryon number, as in a hadronic gas. The
$\langle B S \rangle$ correlator can be expressed in terms of quark
susceptibilites, where the presence of $q\bar q$ bound states should
show up as an enhancement of flavor-off-diagonal elements. Existing
lattice calculations rule these out with great precision. Karsch 
and collaborators have done a similar study for electric charge 
correlations, which are sensitive to the presence of quasiparticles
carrying diquark quantum numbers. Again, the lattice results indicate
a rapid transition from a hadron resonance gas to a perturbative 
quark-gluon plasma in the range $T_c<T<1.4T_c$.

While it may be too early to proclaim the demise of the ``bound state
quark-gluon plasma'' model, these new results are not encouraging
for the model. It is true that the lattice results mentioned above
have nothing to say about gluonic bound states. It would be worthwhile
to explore generalizations of these tests to gluons. A direct application
is impossible, because one cannot introduce a gluonic chemical potential,
but maybe it would be possible to define a chemical saturation factor
$\gamma_g$, similar to the strangeness saturation factor $\gamma_s$ 
which is introduced in hadro-chemical equilibrium models. One would
never have to do a lattice calculation with $\gamma_g \neq 1$, the
parameter would only be used to define derivatives for infinitesimal
deviations from equilibrium. Anyway, model builders will have to abide
by the constraints introduced by these and future lattice results. 

Quite generally, it appears to me that a low viscosity (relative to the
entropy density) of the quark-gluon plasma argues against structure 
formation, except possibly at the nearest neighbor level. Materials 
with long-distance particle correlations, such as polymers, generally 
have large viscosity. The reason is that, although transport cross 
sections can be huge, momentum transfer in these materials is not 
dominated by particle transport, but by the transport of momentum 
along molecular chains or via collective modes. The mean free path for
momentum exchange is therefore large. In the extreme case of crystals,
momentum transfer is caused by phonons, which may travel freely over
extremely large distances, even if the atoms are immobilized. A small
viscosity is thus indicative of a liquid without strong correlations,
except possibly on the scale of the interparticle separation.

Is there any approach, other than euclidean lattice simulations, 
which would allows us to explore the strongly coupled quark-gluon 
plasma? The experience from electromagnetic plasmas is that molecular
dynamics might be a promising method. Two speakers at this conference,
E.~Shuryak \cite{Shuryak} and P.~Hartmann \cite{Hartmann}, presented 
first results from what I would
call ``QCD inspired'' molecular dynamics calculations. As intriguing
as their results are, the fact that these models are constructed in 
the framework of direct two-particle interactions makes them suspect, 
for several reasons: Such models make it very difficult, if 
not impossible, to implement local gauge invariance; also, the 
perturbatively unscreened sector in high-temperature QCD is the 
chromomagnetic sector, which is not easily encoded in two-body forces. 

It is perhaps worth mentioning that a numerical approach to parton 
molecular dynamics already exists, which encodes hard thermal loops 
in Minkowski space via a coupled system of colored partons and a 
spatial lattice \cite{HM-MHM}. In this formulation, which was put to 
use in the work on color instabilities presented here by Nara \cite{Nara}, 
the partons propagate according to Wong's equation
\begin{equation}
\frac{dp_i^\mu}{d\tau} = g Q_i^a F^{a\mu\nu} v_{\nu},
\qquad
\frac{dQ_i^a}{d\tau} = g f_{abc} A^{b\mu} Q_i^c v_{i\mu} ,
\end{equation}
while the soft classical fields satisfy the Yang-Mills equation with 
the particles as sources:
\begin{equation}
\partial_\mu F^{a\mu\nu} = \int d\tau \sum_i g Q_i^a v_i^\nu
                           \delta\left(x-\xi_i(\tau)\right) .
\end{equation}
At weak coupling, this system of equations (first proposed by Heinz
\cite{Heinz}) is known \cite{Kelly} to reproduce the HTL effective theory, 
and this has also been demonstrated numerically \cite{HM-MHM}.
No one has explored what these equations predict for strong coupling. 
Although it is clear that the Heinz model does not reproduce full QCD
when the coupling becomes strong, the fact that we know and understand 
its weak coupling limit provides confidence.
Note that the Heinz model only contains
a mean field and no two-body forces, which may become increasingly 
important at stronger coupling. It would be straightforward and logical 
to include such short-distance interactions as scatterings involving 
color exchange among partons contained in the same lattice cell \cite{BMP}.

An important theoretical playground for those interested in strongly
coupled gauge theories are the supersymmetric gauge theories, for
which a dual gravity-like theory exists. A.~Starinets \cite{Starinets}
reported about the progress in calculating corrections to the results
obtained in the extreme strong coupling limit. For strongly coupled
$N=4$ supersymmetric Yang-Mills theory we now know that the asymptotic
value of the shear viscosity / entropy density ratio is approached as
($N_c$ denotes the number of colors)
\begin{equation}
\frac{\eta}{s} = \frac{1}{4\pi} 
     \left[1 + \frac{135\,\zeta(3)}{(8g^2N_c)^{3/2}} + \cdots\right] .
\end{equation}
Note that $\eta/s$ does not grow for very strong coupling in
this theory, presumably because there is no formation of long-range
structure. Unfortunately, the mapping between the quasiparticles
at weak coupling and those at strong coupling is not well understood
in these SUSY models, and it thus remains unclear to what extent these 
beautiful results apply to real QCD. However, if it were possible to
apply molecular dynamics methods developed for strongly coupled QCD
to the SUSY theories, the analytical results could provide rigorous
tests for the numerical methods.

\section{PHENOMENOLOGY}

\subsection{Electromagnetic probes}

There is little to add to C.~Gale's beautiful review of electromagnetic
probes of dense matter \cite{Gale}. The amazing data presented at this 
conference hold promise for more to come. The di-lepton data from NA60
\cite{NA60} demonstrate that excellent resolution and high statistics are
crucial prerequisites for progress. The 
invariant mass spectra of di-muons from In+In shown by NA60 exhibit not
only a broad source of prompt di-leptons, but even show the narrower 
peak from $\rho$-meson decays in vacuum. Future calculations will thus
have to describe the magnitude of this peak. The data are good enough to
discriminate between different models. Even a 
superficial glance suggests that any model, which predicts only a mass
shift in medium, without broadening, will face great difficulties.

It would be desirable, if theorists would not only compare the new data
with various specific models of the in-medium modifications of the 
$\rho$-meson. Obviously, these modifications can be broken down into
the contributions of various hadronic channels contributing to the
changes in the spectral function of the $\rho^0$. As interesting as 
the obtained insights are, they may obscure the answer to more general 
questions, such as: What is the route to chiral symmetry restoration and
deconfinement? Is it the general broadening of hadronic states until
they dissolve into a continuum? Or does the transition occur primarily 
through shifts in hadron masses or mass splittings between chiral 
partners? It would be useful to interpret the di-lepton data in terms
of such deeper concepts in a model independent way.

\subsection{Jet quenching}

We have seen a lot of (by now) routine phenomenology on jet quenching, 
but also some interesting new ideas. N.~Borghini \cite{Borghini} presented 
an approach, in which the medium effects are simply encoded in a 
one-parameter change of the splitting function within the well 
established modified leading logarithmic approximation (MLLA) theory of 
jet fragmentation. The resulting $dN/dz$ distribution of jet fragments 
looks just like what is observed in Au+Au collisions when the threshold
for back-to-back coincidences is set low. It will be interesting to see
how far this idea can be pushed and whether the new parameter in the
splitting function can be microscopically calculated. Another promising
new development is the theory of di-hadron fragmentation functions,
discussed by A.~Majumder \cite{Majumder}. 

There is a sense among theorists
now -- especially after the Cu+Cu data did not generate a new surprise -- 
that jet quenching at RHIC is sufficiently well understood to justify
detailed predictions for the LHC, as presented in C.~Salgado's talk
\cite{Salgado}. Such efforts should be encouraged, because a detailed set 
of prediction will allow us to assess the quality and predictiveness of 
the theory when the LHC data arrive in a few years.

But let me inject a word of caution. As first demonstrated by Armesto 
et al. \cite{jets+flow}, collective flow can influence the energy loss
of a parton in the quark-gluon plasma. This is especially true, if 
the longitudinal expansion is not boost invariant, leading to a modified
density profile with time. As Renk and Ruppert recently showed \cite{RR05}, 
the deduced energy loss parameter $\hat q$ can differ by a factor of
five (!) between a boost invariant scenario and one with longitudinal
acceleration and strong transverse flow. The fact that the latter would
give a much larger suppression for the same value of $\hat q$ could
help counteract, e.~g., the effect of a larger initial quark-to-gluon 
ratio as suggested by Lappi's calculations \cite{Lappi} mentioned above.
As the phenomenology of jet quenching becomes increasingly quantitative, 
it will be important to compare different calculations of parton energy 
loss for the same space-time evolution model.

On a personal note, I found the results on single hadron suppression
($R_{AA}$) and additional back-to-back suppression ($I_{AA}$) shown
by the STAR collaboration \cite{Jacobs}, especially exciting. For 
the specific cuts set by STAR, the two factors were equal: $R_{AA}
\approx I_{AA} \approx 0.23$. Why should this be the case? As I and
others have argued, $R_{AA}$ is a geometric suppression factor which 
represents the size of the surface layer from which observed leading 
hadrons are emitted, compared with the total volume in which jets are 
produced. Similarly, $I_{AA}$ can be viewed as an additional reduction factor 
describing the width of the ``equatorial'' torus from which the observed 
di-jets are tangentially emitted with an approximately equal and modest 
energy loss (see Figure). I predicted this relationship in a somewhat
schematic treatment of jet quenching a few years ago \cite{BM03} and
am happy to see the prediction substantiated. Clearly, the relative
value of $R_{AA}$ and $I_{AA}$ will depend on the selected momentum
thresholds, but the general argument suggests that the connection 
could be another tool in the analysis of dense matter effects on hard
partons.

Some of the lectures at this meeting, experimental as well as 
theoretical ones (see e.~g. \cite{Cole,XNWang}, have conveyed the 
message that jet quenching in Au+Au and Cu+Cu collisions at RHIC is 
quantitatively well understood. How sure are we of this? Besides the
uncertainties already mentioned (influence of flow, gluon dominance
of the early quark-gluon plasma), there is a glaring discrepancy 
between different approaches, which has not been resolved. Those
who use the data to deduce an effective value of the energy loss
coefficient $\hat q = 5\cdots 15\, {\rm GeV}^2/{\rm fm}$ \cite{Dainese}
(effective, because the quoted value approximately corrects for 
longitudinal expansion) have noted that this value is significantly 
larger, by a factor five or so, than the value predicted years ago 
by Baier \cite{Baier} on the basis of perturbative QCD. On the other 
hand, those who apply perturbative calculations of radiative energy 
loss to calculate $R_{AA}$ in nuclear collisions \cite{XNWang,Vitev}
find good agreement with the data. 

There are two possible resolutions to this conundrum. One would be
that the standard fits of $\hat q$ overestimate the value of this
parameter due to unrealistic assumptions, e.~g.\ about the space-time
evolution of the energy density, as discussed above. 
Another one would be that Baier's
estimate of $\hat q$ as a function of the energy density represents
a serious underestimate of the true effective value of $\hat q$ in 
perturbative QCD. I believe that both sides owe us a careful analysis 
of the problem, on which everyone can agree. Because jet quenching is
such a central aspect of the new discoveries by the RHIC physics 
program, it is important that the apparent contradiction be resolved
soon. 

Another challenge for theorists is, of course, the development of 
techniques for a nonperturbative calculation of $\hat q$ on the lattice.
This does not seem totally hopeless, because (a) gluon dominance makes
quenched lattice simulations suitable for this property of the medium,
and (b) the parameter $\hat q$ can be formulated in terms of correlators
along the light cone \cite{Kovner} of the form
\begin{equation}
\int_0^{x_-} dx'_- \, \left\langle F_{+i}(x_-) W(x_-,x'_-) F^{+i}(x'_-)
        W^\dag(x_-,x'_-) \right\rangle , 
\end{equation}
which could perhaps be calculated by means of analytic continuation 
using maximum entropy techniques. Even if this turns out to be a very 
hard computational problem, it needs to be attacked, because the claim 
that the quark-gluon plasma seen at RHIC is strongly coupled and 
nonperturbative in view of its low viscosity and the claim that 
perturbative QCD describes the data on jet quenching are not easily 
reconciled. At the very least, it would be important to get a 
quantitative understanding of the relevant QCD scale for $\hat q$,
in order get better control of its value in perturbation theory.
This would require a next-to-leading order calculation of radiative
energy loss in the medium.

\subsection{Charmonium}

Quite possibly the most challenging set of data for theorists, which
was first presented at this conference, are the PHENIX results on
charmonium production in Au+Au and Cu+Cu. Even with their limited 
statistics, they rule out any simple extrapolation of the models of 
comover suppression developed to explain the CERN-SPS data. Two 
diametrically opposed conclusions have been drawn from the data,
as discussed by M.~Nardi \cite{Nardi}: one possibility is that the
$J/\Psi$ is not ``anomalously'' suppressed, both at the SPS and RHIC,
the other one is that there is a fair amount of regneration of the
$J/\Psi$ at RHIC by $c-{\bar c}$ recombination. The first alternative
is supported by the recent results of quenched lattice calculations
of the $c-{\bar c}$ spectral functions, which show the survival of 
a pronounced resonance in the vector meson channel up to $T>1.5T_c$
\cite{J/Psi-SF}. The second alternative is supported by the growing 
evidence from the RHIC experiments that $c$-quarks participate in the
collective flow, suggesting that they are also subject to statistical
hadronization by coalescence. Finally, there is the possibility that
what has been called ``anomalous'' suppression at SPS energies is 
caused by the absorption of excited charmonium states, especially the
$\chi_c$, feeding into the observed $J/\Psi$ yield. If these states 
are completely absorbed in the most central events at the SPS, then
no additional suppression due to this mechanism can occur at RHIC.

The problem, in my view, is that we still do not have a comprehensive
and generally accepted theoretical framework, in which to discuss
$J/\Psi$ suppression. The reason for this lack is not clear. If 
the $J/\Psi$ acts as a ``hard'' probe, it should be possible to derive 
a systematic formulation of its production and destruction in heavy
ion collisions within the framework of perturbative QCD and effective
field theory (e.~g., nonrelativistic QCD). The heavy quark mass 
provides a large scale, which can serve to separate nonperturbative
physics of the dense medium from the dynamics of the heavy quark. 
Such a formulation might only be marginally valid for $c$-quarks, 
but it would surely work well for $b$-quarks, which will soon become
the center of attention at the LHC. On the other hand, if such a
formalism cannot be found, we will need to give up on the notion 
that the $J/\Psi$ (or the $\Upsilon$!) is a ``hard'' probe. The
ball is clearly in the theorists' court, and the new data provide
ample motivation to address this issue vigorously.

On specific issues, the failure of direct lattice calculations of
the $c-{\bar c}$ spectral functions \cite{Petreczky} and potential 
models motivated by the lattice results \cite{Mocsy} is puzzling.
The spectral functions derived from the potential models do not 
agree with those obtained by analytic continuation directly from 
the lattice. I am not sure we fully understand the reason for this 
discrepancy. The spectral functions derived from the lattice do not
indicate the presence of a continuum in the vector meson channel.
Because the $D-{\bar D}$ channel can be easily incorporated into 
potential models, but remains absent in quenched lattice calculations 
of the spectral functions, the role of dynamical quarks in the 
survival of heavy quarkonium states at $T>T_c$ remains an issue of 
concern. It is unclear, however, why the lattice results do not 
reveal a gluonic continuum at $T \geq 1.5\, T_c$. 

The existence of a possibly broadened peak in the $J/\Psi$ spectral 
function does not necessarily mean that the initially created $J/\Psi$
mesons survive in the medium. The present lattice calculations may not 
reflect the true width of the state and thus may not give a good estimate 
of its lifetime inside the quark-gluon plasma. On the topic of 
regeneration, more detailed calculations of $J/\Psi$ formation by 
$c-{\bar c}$ recombination are urgently needed. Given the success 
of recombination models for light hadrons, reliable calculations 
should be within reach. Overall, it is clear that there is much room 
for theoretical improvements in the area of charmonium suppression.

\subsection{Hadronization}

As W.~Florkowski's \cite{Florkowski} talk aptly demonstrated, the 
dispute about the
relationship between chemical and kinetic (``thermal'') freeze-out
of hadrons at the end of a heavy ion collision remains unresolved. 
Clearly, there exist strong theoretical arguments for a differential
freeze-out of various hadrons depending on their mean free paths 
in an equilibrated hadron gas. Hadrons with no or few known resonant
excited states, such as the $\phi$-meson or the $\Omega$-hyperon 
should freeze out early. On the other hand, a common general freeze-out
of hadron abundances and momentum spectra would require a specific 
mechanism forcing a sudden termination of efficient equilibration 
mechanisms. If such a scenario were realized in nature, I can only
imagine it to be associated with the hadronization transition itself.
The fact that, according to lattice QCD, this transition is smooth 
and not discontinuous in thermal equilibrium, such a scenario will
require strongly off-equilibrium conditions during hadronization (such 
as the sudden decay of a deeply supercooled state \cite{Csorgo,RL00}).

One can ask whether the available data already provide conclusive 
evidence for or against a common freeze-out. The new high resolution
di-lepton data from NA60 should be sufficient to set a lower bound 
on the duration of the strongly interacting hadron gas phase. The idea
that the $\rho$-meson yield seen through the lepton-pair decay channel
measures the lifetime of the hadronic phase is not new \cite{Heinz-rho}.
But only now do we have data from the SPS, which tell us how many 
$\rho^0$'s decay inside a hadronic medium, which is dense enough to
modify the spectral function. This broadening is related to the rate
of binary collisions which, in turn, maintain the kinetic (but not 
necessarily the chemical) equilibrium of the hadronic medium. Single
freeze-out models are thus challenged to describe the observed yield 
of low-mass di-lepton emission. Other data, such as the results from 
STAR on the yield of strange resonances as function of centrality 
(see S.~Salur's talk \cite{Salur}) should also provide serious tests 
for freeze-out models. The proponents of these models need to address
these results.

Let me now turn to strangeness. Most fits to the data yield an almost 
perfect abundance equilibration of strange hadrons ($\gamma_s=1$), 
which was predicted long ago as one of the characteristic signatures 
of deconfinement and chiral symmetry restoration. However, there exist 
other high-quality fits of hadron yields \cite{LR05}, which invoke a 
significant oversaturation of strangeness. Furthermore, because of the 
different masses of the carriers of strangeness in the quark-gluon 
plasma and the hadron gas phase, the value of $\gamma_s$ does not 
have to remain the same during the transition. This suggests that it
might be an interesting intellectual exercise to study the equation 
of state of QCD as a function of $\gamma_s$, the overall strangeness
abundance. As in the case of $\mu_B$, one would probably want to 
calculate the derivatives with respect to $\ln(T\gamma_s)$ at the 
equilibrium point $\gamma_s=1$. In many respects, this method would
correspond to studying the effects of a varying number of flavors
on the equation of state, which we know to be nontrivial.

The recombination model for hadron production at intermediate transverse
momenta continues to work amazingly well. The extended range of particle
identification in Runs 4 and 5 is finally allowing the data to probe the
predicted transition between the region dominated by recombination and 
the high-momentum region dominated by fragmentation. Results shown
at this conference \cite{Barannikova} confirm the theoretical predictions 
of a transition between 4 and 7 GeV/c, and theorists are eagerly waiting 
for more analyzed data. 

How quickly the experiments move to challenge theoretical predictions 
was demonstrated by STAR's presentation of data on the systematic 
baryon/meson violations of the valence quark number scaling of the 
$p_T$-dependence of the elliptic flow. Including higher Fock states 
in the hadron wave functions, we had recently predicted deviations at 
the few-percent level \cite{MFB05}. A second source of such deviations
from the scaling law can be caused by the internal momentum distribution
of constituent quarks \cite{Greco04}. How exciting to see that the data 
confirm the existence of such deviations with the correct sign 
\cite{Sorensen}! More work will need to be done to determine whether
the deviations from universality provide another confirmation of the 
recombination model, or whether they constitute a challenge. Other 
theoretical progress reported at this conference were the effects of 
parton correlations on same-side di-hadron yields \cite{Bass} and the 
possible origin of the pedestal effect in the same data \cite{Hwa}.

\subsection{Fluctuations and Correlations}

As R.~Lednicky's talk \cite{Lednicky} made clear, the ``HBT puzzle'' of 
the RHIC data is still with us. Generally the measured correlations indicate 
a short overall duration of the reaction, strong transverse flow, and a very 
brief, almost sudden freeze-out. Theory still has great trouble getting
the small value $R_{\rm out}/R_{\rm side}\approx 1$ right. As the success 
of hydro-motivated parametrizations of the freeze-out shows 
\cite{HBT-RL,HBT-CCLS}, it is not impossible to reproduce the data. 
The problem lies in finding a dynamical evolution model, which yields 
the right freeze-out parameters. There is justified hope that the new 
3-D hydro plus cascade codes may change the situation. Stay tuned to 
see whether this venerable problem finds a resolution soon.

\section{CONCLUSIONS}

This conference has been a showcase of significant progress in the 
theory of superdense hadronic matter and its formation and evolution 
in relativistic heavy ion collisions. It is certainly a very productive
time for theorists in our field, because the experiments provide for a 
wealth of data which require interpretation and explanation. The data
also make it easier than ever before to verify or repudiate theoretical
approaches and models. As I have tried to point out, the successes of
theory are not evenly distributed. There are some areas, where theory is
well on its way toward definite descriptions, and others, where even a
comprehensive framework is still lacking and it is time to catch up. 
Above all, however, it is upon us theorists to embrace the offer made
by one of the earlier speakers in this session \cite{Akiba}: 
``We look forward to working  with the theory community to extract the 
properties of the matter (produced in relativistic heavy ion collisions).''

\section{ADDITIONAL COMMENTS}

The following remarks were motivated by comments received after posting
a draft version of my summary talk. 

\subsection{QGP or ``sQGP''?}

The characterization of the matter produced in nuclear collisions at
RHIC is an experimental issue. On the other hand, the predictions of
QCD for the properties of thermal QCD matter in the temperature range 
$T_c \leq T \leq 2T_c$ are a theoretical issue. As discussed above, 
it has been widely argued that the quark-gluon plasma in this temperature 
range is a strongly coupled state of matter -- a ``sQGP'' 
\cite{Gyulassy:2004,Shuryak}. How compelling are the theoretical 
arguments for this claim? 

It is possible to derive an effective theory for the long-distance 
dynamics (momenta $k < T$) of thermal gauge theories. If the dynamics
at momentum scales $k \geq T$ is weakly coupled, the effective theory 
can be derived perturbatively. Two forms of this effective theory are
known. One is the hard-thermal loop (HTL) effective theory \cite{HTL},
the other is the dimensionally reduced theory based on local interactions
of the gauge field \cite{Kajantie,Braaten}. In both frameworks one can
ask how strong the effective coupling $\alpha_s^{\rm eff}=g^2/4\pi$ is 
as a function of temperature. This question has recently been studied 
by Laine and Schr\"oder \cite{Laine}, who calculated the effective 
action for QCD in the dimensionally reduced theory up to two-loops. 
They found that the effective coupling constant $\alpha_s =g_E^2/4\pi$ 
is well controlled down to $T=T_c$ and remains remarkably small: 
$\alpha_s^{\rm eff}(T_c) \approx 0.28$. These authors also found that
the effective theory agrees very well with full thermal QCD for the 
spatial string tension, which is an important measure of the long
distance dynamics. Similar conclusions are reached in the framework of 
the HTL approach \cite{ABS} which, however, has not yet been carried out 
to full two-loop order and thus does not yet permit the extrapolation 
all the way down to $T_c$.

These studies suggest that the quark-gluon plasma in the temperature 
region near $T_c$ is well described by an effective theory with a 
modestly strong coupling constant. This raises the question, where 
the transition between the weak coupling and strong coupling regimes
of thermal QCD occurs: at $T_c$ or at some $T>T_c$? This cannot be 
decided by numerology, because the perturbative expansion parameter
is, generally, $\alpha_sN_c/\pi \approx \alpha_s$, while the expansion 
parameter of the strong coupling expansion of string duals of nonabelian
gauge theories is $\lambda=g^2N_c$ \cite{Klebanov}, and the conditions 
$\alpha_s\ll 1$ and $\lambda\gg 1$ are both fulfilled in the vicinity
of $T_c$. Turning to physical arguments, it is noteworthy that the 
perturbative quasiparticles of the effective theory, thermal gluons and 
plasmons, are short-lived. The collisional width of a gluon/plasmon 
at rest is given by \cite{BP90} 
\begin{equation}
\Gamma_g(0) \approx \frac{1}{2\pi}g^2N_c\, T \approx 1.5\, T .
\end{equation}
On the other hand, the effective mass of a gluon/plasmon at rest is 
$m_g^* = gT\sqrt{N_c/9} \approx T$. Roughly the same relationship holds 
for thermal quarks \cite{BP92}. In other words, all slowly moving 
quasiparticles are strongly damped. The characteristic nature of the 
temperature region near $T_c$ may thus be that the quasiparticles, 
gluons and quarks above $T_c$ and hadrons (such as the $\rho$-meson) 
below $T_c$, are strongly collision broadened. Such a property would 
be indicative of a liquid, which is characterized by the absence of 
long-lived quasiparticles and long-range order. Because the effective
coupling $g_E$ changes only slowly with temperature, it is by no means
clear how for above $T_c$ one needs to go before the widths of the
plasma quasiparticles become a higher-order effect compared to their
masses.

\subsection{Viscosity}

Does the apparent nearly ``perfect fluid'' property of the matter 
observed at RHIC ($\eta/s \ll 1$) imply that the quark-gluon plasma 
has, by some measure, a very low shear viscosity or is it simply a
signature of a large number of degrees of freedom, i.~e.\ a large
entropy density, as Hirano and Gyulassy \cite{Hirano} have recently 
argued? The general expression for the shear viscosity is
$\eta \sim \rho{\bar p}\lambda_f$, 
where $\rho$ is the particle density, $\bar p$ the average momentum 
of a partice, and $\lambda_f$ the mean free path. Since 
$\lambda_f=(\sigma_T\rho)^{-1}$, where $\sigma_T$ denotes the 
transport cross section, the density cancels from the expression, 
and we have 
$\eta \sim {\bar p}/\sigma_T$.
Because cross sections are unitarity bounded ($\sigma_T\leq 4\pi/{\bar p}^2$
for s-wave scattering), we find approximately
$$\eta \geq {\bar p}^3/4\pi \sim 27\, T^3/4\pi , $$
where we made use of the relation ${\bar p}\approx 3T$ for 
relativistic matter. It is against this value that the smallness of 
the viscosity of the quark-gluon plasma should be judged.

On the other hand, the entropy density of a relativistic medium is
approximately $s \sim 4\rho$, allowing us to write 
$\eta/s \sim {\bar p}\lambda_f/4$. 
The smallness of $\eta/s$ is thus a measure of the smallness of 
the mean free path, compared with the thermal wavelength 
$\lambda_{\rm th} \sim 2\pi/{\bar p}$ of a particle. As such, it is
not directly sensitive to the number of degrees of freedom in the
medium.

\section*{Acknowledgements}
{\small This work was supported in part by the U.\ S.\ Department
of Energy under grant DE-FG02-05ER41367.} My special thanks go to the
organizers of this conference for making it such a stimulating and
productive experience. I am grateful for comments from M.~Asakawa,
T.~Cs\"org\"o, I.~Dremin, D.~d'Enterria, T.~Hirano, C.M.~Ko, 
S.~Mrowczy{\'n}ski, P.~Petreczky, R.~Pisarski, R.~Rapp, S.~Wong, 
and W.~Zajc on an earlier version of this manuscript. I also thank 
M.~Strickland for helpful communications.

\end{document}